\documentclass[a4paper,11pt]{article}
\usepackage{pos}
\usepackage{subcaption}

\title{First results from Hybrid Hadronization in small and large systems}
\ShortTitle{Hybrid Hadronization}

\manuallySeparateAuthors

\author*[a]{Michael Kordell}

\author{ for the JETSCAPE collaboration}

\affiliation[a]{Cyclotron Institute, Texas A\&M University,\\
  College Station TX 77843, USA}

\emailAdd{mkordell@tamu.edu}

\abstract{
“Hybrid Hadronization” is a new Monte Carlo package to hadronize systems of partons. It smoothly combines quark recombination applicable when distances between partons in phase space are small, and string fragmentation appropriate for dilute parton systems, following the picture outlined by Han et al. [PRC 93, 045207 (2016)]. Hybrid Hadronization integrates with PYTHIA 8 and can be applied to a variety of systems from $e^++e^-$ to $A+A$ collisions. It takes systems of partons and their color flow information, for example from a Monte Carlo parton shower generator, as input. In addition, if for $A+A$ collisions a thermal background medium is provided, the package allows sampling thermal partons that contribute to hadronization. Hybrid Hadronization is available for use as a standalone code and is also part of JETSCAPE since the 2.0 release.\\
In these proceedings we review the physics concepts underlying Hybrid Hadronization and demonstrate how users can use the code with various parton shower Monte Carlos. We present calculations of hadron chemistry and fragmentation functions in small and large systems when Hybrid Hadronization is combined with parton shower Monte Carlos MATTER and LBT. In particular, we discuss observable effects of the recombination of shower partons with thermal partons.
}

\FullConference{%
  HardProbes2020\\
  1-6 June 2020\\
  Austin, Texas}

\begin{document}
\maketitle

In high energy collisions, from dilute systems such as in $e^+ + e^-$ events to the dense systems in $A+A$ events, the observed particles typically include some number of hadrons.  The formation of these hadrons in hadronization is not well understood from first principles.  However, there are hadronization models that can accurately describe observables related to hadron production.  The models pertinent to this study are Lund string fragmentation \cite{Andersson:1983ia} as implemented in the PYTHIA 8 \cite{Sjostrand:2014zea} event generator and quark coalescence / recombination \cite{Fries:2008hs}.

Lund string fragmention builds from the property of color flux tube formation in the QCD vacuum since color flux is expelled. This gives string-like behavior with gluons forming kinks in these strings connecting color charges at large distances.  These strings are then fragmented to form hadrons.  String fragmentation enjoys a great deal of success in describing hadronic observables for dilute systems formed in $e^+ + e^-$ and $p+p$ collisions.

On the other hand, in a densely populated parton system quarks and quark-like constituents could directly recombine into hadron and hadron resonances, similar to recombination in atomic physics.  Quark recombination can describe several key observables in $A+A$ collisions such as enhanced baryon production and elliptic flow scaling.

If we consider jet production in $A+A$ events, we have a system composed of both dilute and dense portions.  For such a case, neither string fragmentation nor quark recombination is expected to be applicable over the entire event.  This is the motivation for \textit{Hybrid Hadronization}.  It is a hybrid of these two hadronization models that extrapolates smoothly between vacuum phenomenology of string fragmentation and recombination in a densely populated environment, with a focus on hadronization of parton showers in these systems.  Specifically, the motivation is to study in-medium effects of jet hadronization such as hadron chemistry, momentum diffusion, and medium flow effects.

The implementation of Hybrid Hadronization was initially developed as part of the JET collaboration and is now part of the JETSCAPE framework \cite{Kauder:2018cdt}.  The algorithm used \cite{Han:2016uhh} is based on the instantaneous recombination model \cite{Fries:2008hs} and begins with some partons from some shower Monte-Carlo.  The gluons present are split into $q \bar{q}$ pairs.  Quarks that are close in coordinate and momentum space have some probability to recombine into hadrons.  Recombination probability is given by the overlap of the quark wavefunctions with the bound state hadron wavefunction and can be calculated using the Wigner formalism assuming a Gaussian wave packet form for the quarks and fitting the hadron wavefunction widths to measured or predicted charge radii.  Holes in the strings produced are naturally repaired using color flow information given from the shower Monte-Carlo.  The remnant strings are then fragmented into hadrons using PYTHIA 8.

In the presence of a medium, this procedure can be extended to include recombination of shower partons with thermal partons.  All partons in the jet that are ready to be hadronized must exist at or outside the surface of the quark gluon plasma (QGP).  If there are partons that exist within the QGP, they must either be propagated to the surface by some shower Monte-Carlo or absorbed by the medium.  Sampled thermal quarks are added to the list of shower quarks to be considered for recombination.  The same recombination procedure is applied and remnant strings are fragmented as before.  As we are only considering jet hadronization effects, we do not include purely thermal hadrons for our analyses.  This must be taken into account for any direct comparisons to experimental data.

\begin{figure}[h!]
\centering
\begin{subfigure}{.475\textwidth}
  \centering
  \includegraphics[width=\linewidth]{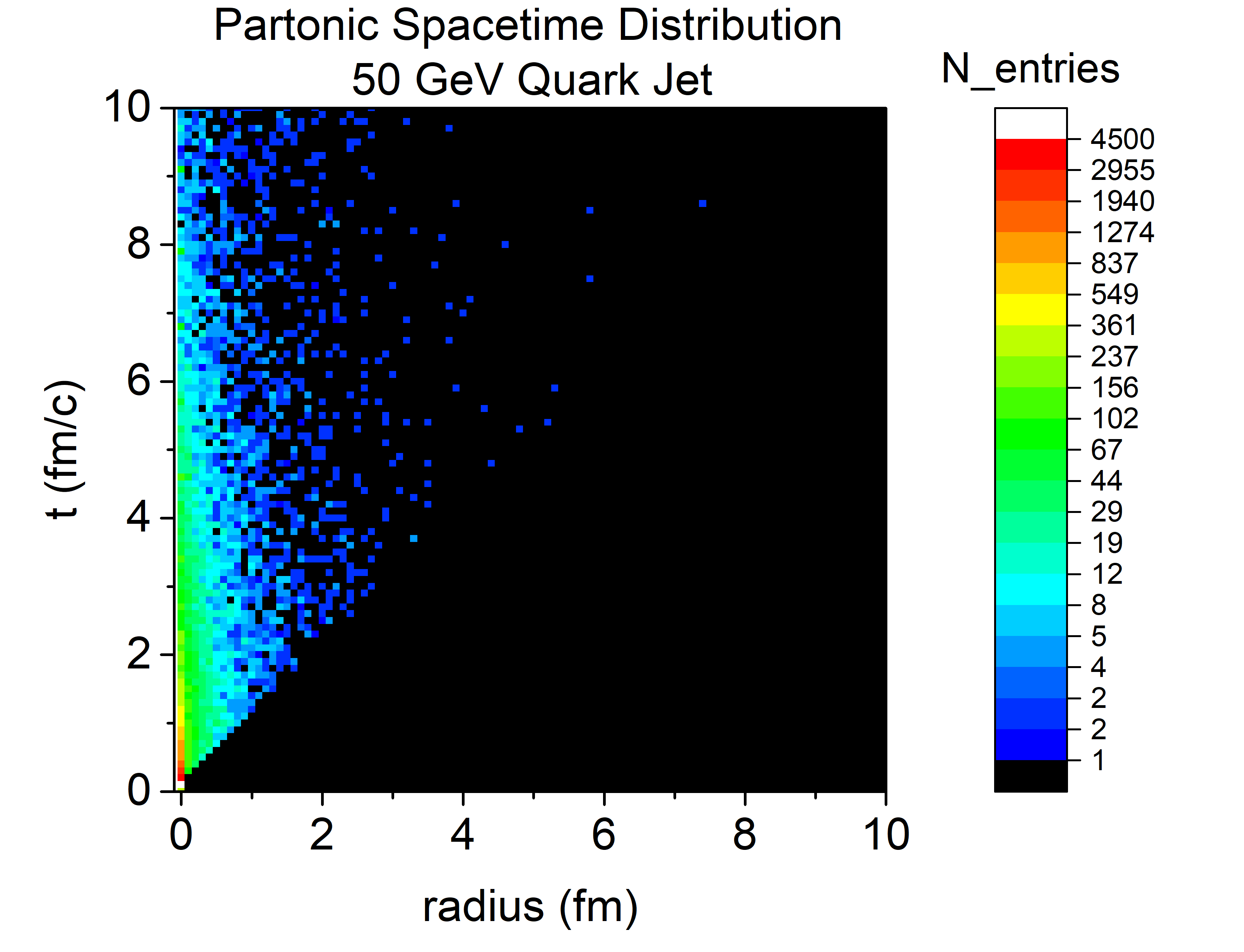}
  \caption{The spacetime distribution of partons ready for hadronization ($Q<Q_0 =$ 1 GeV) in a 50 GeV quark jet in vacuum.}
  \label{fig:fig1b}
\end{subfigure}
\hfill
\begin{subfigure}{.475\textwidth}
  \centering
\vspace{-0.175in}
  \includegraphics[width=\linewidth]{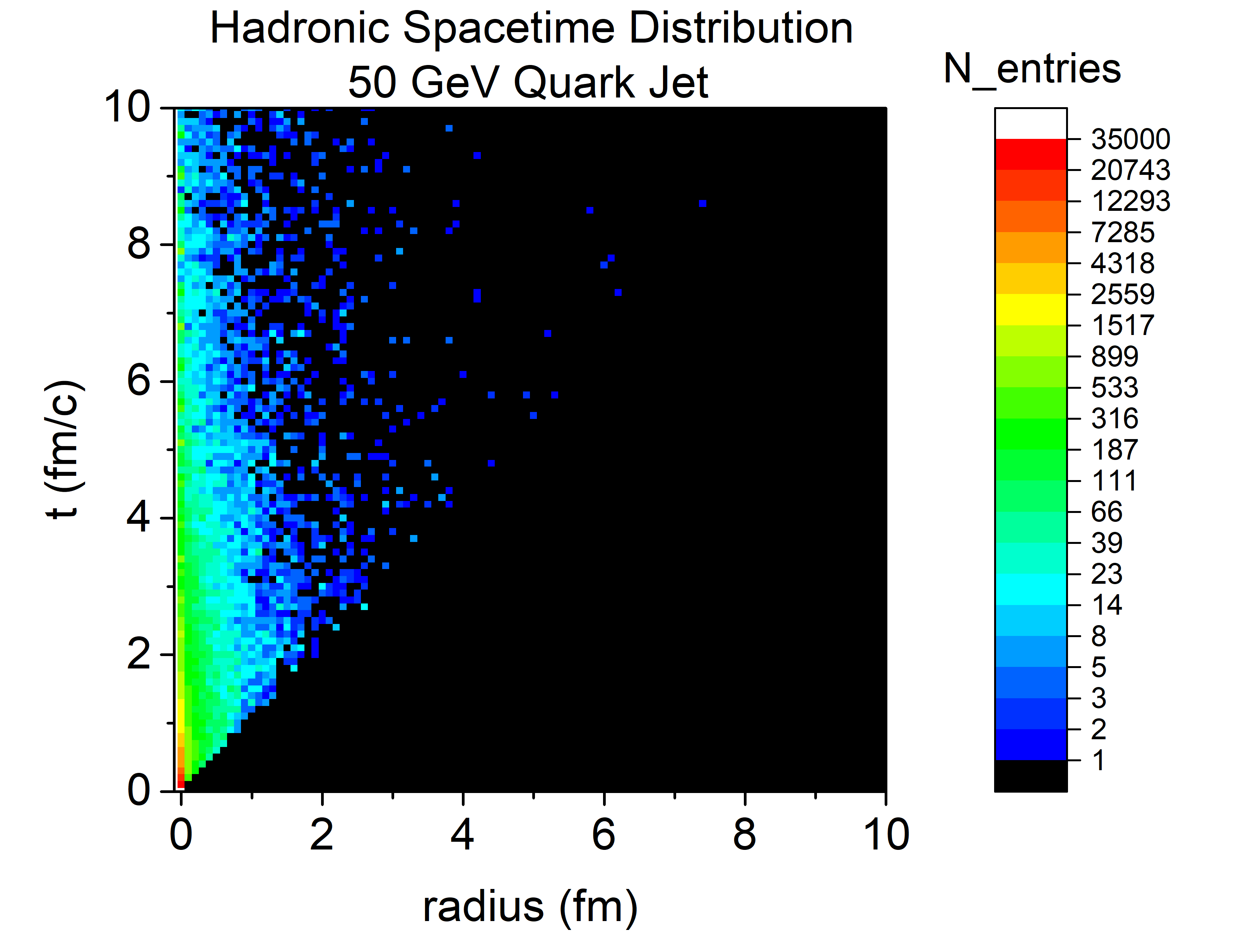}
  \caption{The spacetime distribution of hadrons immediately after hadronization in a 50 GeV quark jet in vacuum.}
  \label{fig:fig1c}
\end{subfigure}
\vskip\baselineskip
\begin{subfigure}{.475\textwidth}
  \centering
  \includegraphics[width=\linewidth]{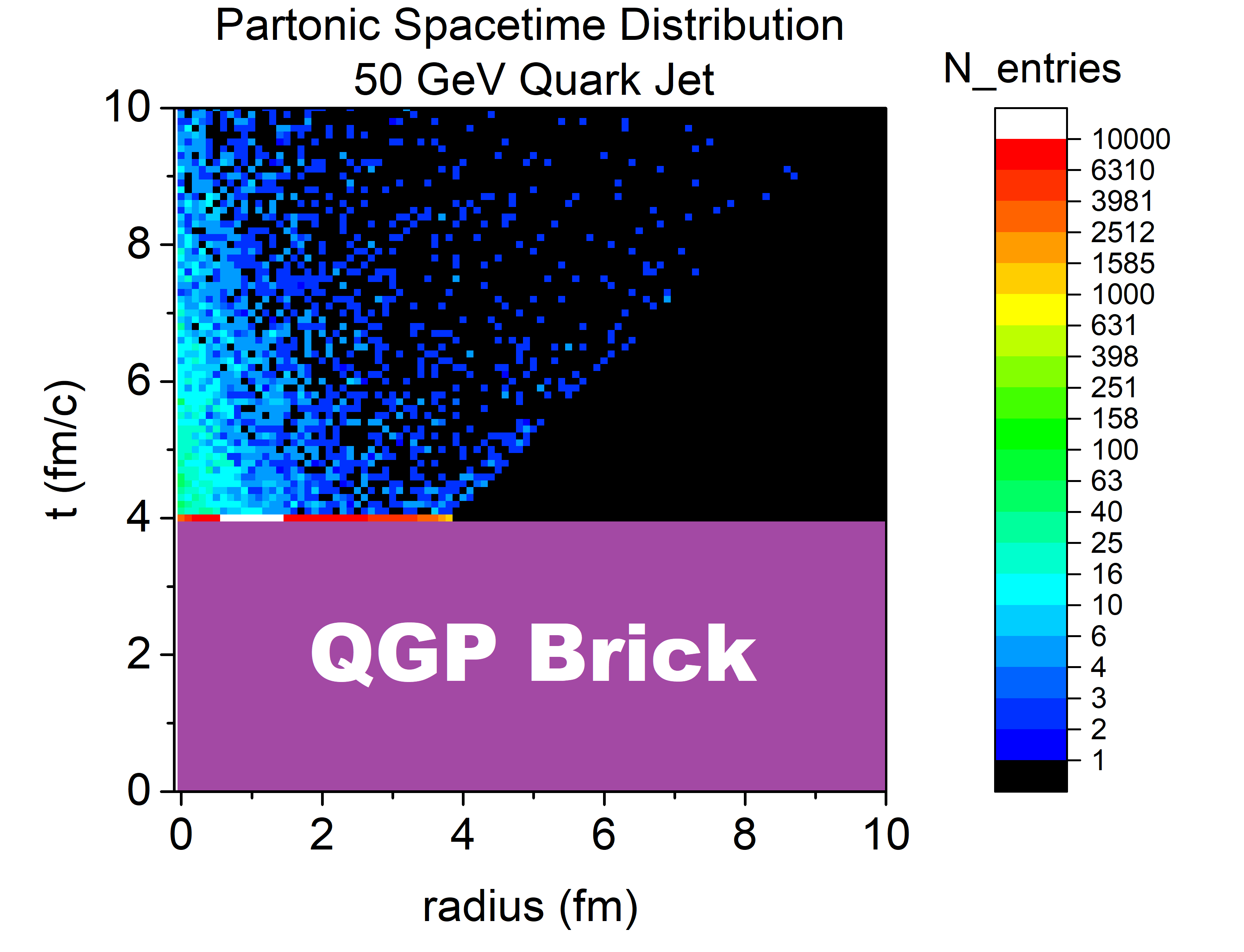}
  \caption{The spacetime distribution of partons ready for hadronization ($Q<Q_0 =$ 1 GeV and $T<T_C$) in a 50 GeV quark jet in a 4 fm brick.}
  \label{fig:fig1d}
\end{subfigure}
\hfill
\begin{subfigure}{.475\textwidth}
  \centering
\vspace{-0.175in}
  \includegraphics[width=\linewidth]{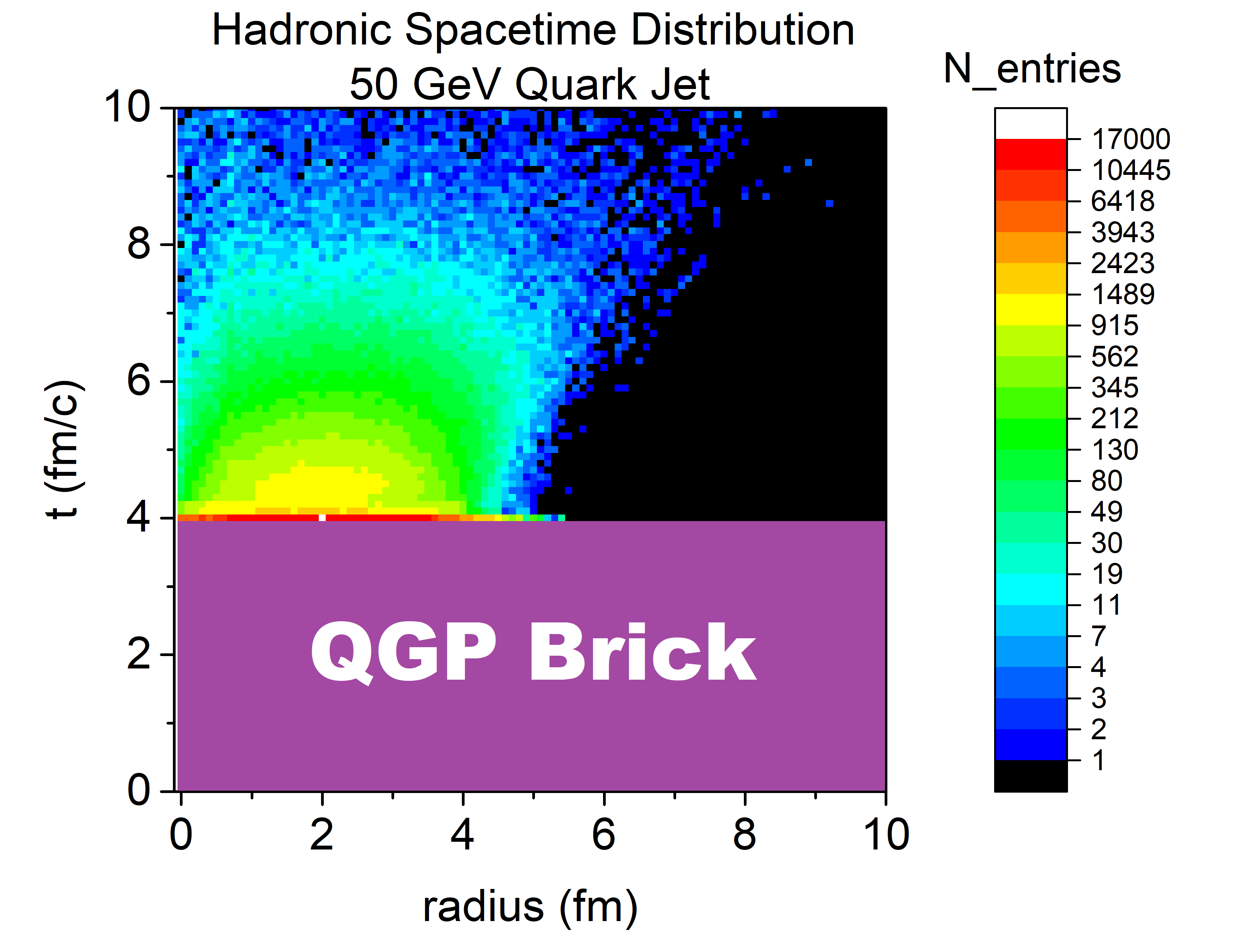}
  \caption{The spacetime distribution of hadrons immediately after hadronization in a 50 GeV quark jet in a 4fm brick.}
  \label{fig:fig1e}
\end{subfigure}
\caption{Spacetime structure for 50 GeV quark jets immedately before and after hadronization.}
\label{fig:fig1}
\end{figure}

The space-time structure of a heavy-ion event can influence a number of observables and is important to hadronization.  The parton shower typically extends further in space-time than the fireball size; for a 500 GeV jet this can be more than 100 fm/$c$.  In Fig.\ \ref{fig:fig1} the space-time structure of 50 GeV quark jets is shown in vacuum and in the presence of a medium for partons that are ready for hadronization and the produced hadrons immediately after hadronization.  In vacuum, the partonic and hadronic space-time structures look similar as to be expected, with the development of the lightcone on the diagonal.  The presence of a medium drastically changes the distribution of hadrons.  At partonic level, the core of the jet punches through the medium and there is an intense disk of jet partons at the hadronization hypersurface.  At hadronic level, the core of the jet protrudes out of the brick and there is a thich halo of hadrons above the hypersurface.  There is a smearing of the hadron position on the order of the hadron size, due to recombination of shower partons with thermal partons that lie outside of the lightcone.

For this study of Hybrid Hadronization, a QGP brick with a space-like hypersurface was considered and the size was varied.  Additionally, the thermal partons sampled had a flow velocity that was varied.  The jet initiating parton is a fixed energy 20 GeV quark, showered with MATTER \cite{Majumder:2013re} \cite{Cao:2017qpx} and LBT \cite{He:2015pra} \cite{Cao:2016gvr}, and hadronized with Hybrid Hadronization.  The traditional recombination signals of an enhanced baryon to meson ratio and flow were studied.  It should be pointed out that these plots do not include purely thermal hadrons as this study is of the systematics of Hybrid Hadronization.  A comparison to experimental data will necessitate the inclusion of these.

\begin{figure}[h!]
\centering
\begin{subfigure}{.475\textwidth}
  \centering
  \includegraphics[width=\linewidth]{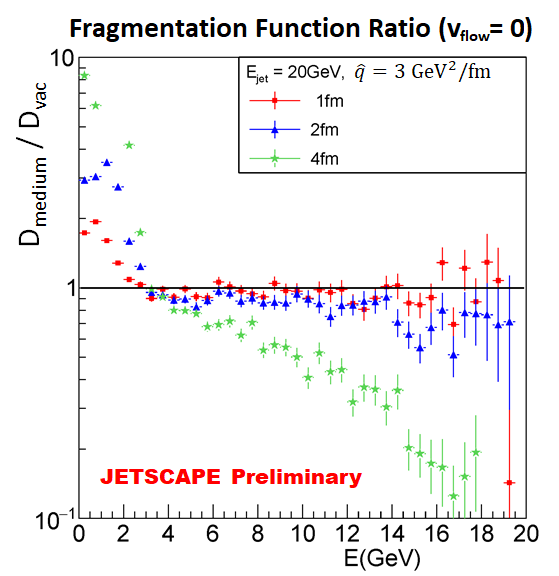}
  \caption{Ratio of fragmentation functions for 20 GeV quark jets where sampled thermal partons were given no flow velocity.}
  \label{fig:fig2a}
\end{subfigure}
\hfill
\begin{subfigure}{.475\textwidth}
  \centering
  \includegraphics[width=\linewidth]{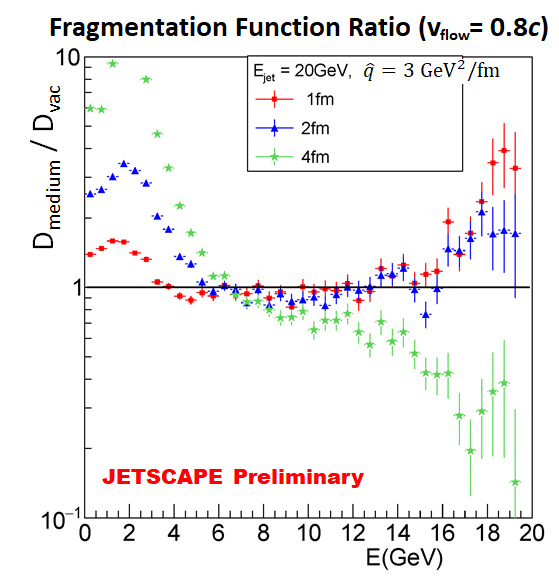}
  \caption{Ratio of fragmentation functions for 20 GeV quark jets where sampled thermal partons were given a flow velocity $v_{flow} = 0.8c$.}
  \label{fig:fig2b}
\end{subfigure}
\caption{Ratio of fragmentation functions for 20 GeV quark jets in bricks of varying sizes to vacuum.}
\label{fig:fig2}
\end{figure}

\begin{figure}[h!]
\centering
\begin{subfigure}{.475\textwidth}
  \centering
  \includegraphics[width=\linewidth]{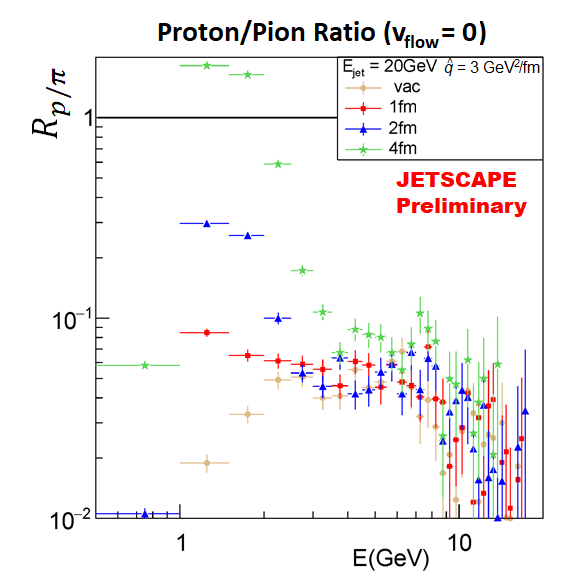}
  \caption{Ratio of proton to pion yield for 20 GeV quark jets where sampled thermal partons were given no flow velocity.}
  \label{fig:fig3a}
\end{subfigure}
\hfill
\begin{subfigure}{.475\textwidth}
  \centering
  \includegraphics[width=\linewidth]{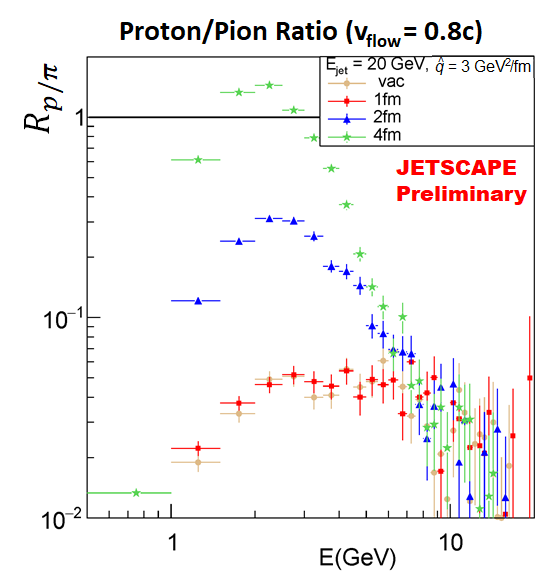}
  \caption{Ratio of proton to pion yield for 20 GeV quark jets where sampled thermal partons were given a flow velocity $v_{flow} = 0.8c$.}
  \label{fig:fig3b}
\end{subfigure}
\caption{Proton to pion ratio for 20 GeV quark jets in bricks of varying sizes and in vacuum.}
\label{fig:fig3}
\end{figure}

In Fig.\ \ref{fig:fig2} the ratio of fragmentation functions for bricks of varying sizes to vacuum show an enhancement at low energy due to recombination with thermal partons.  This enhancement is pushed out to a higher energy if the sampled thermal partons have a collective flow velocity.  This enhancement increases with the size of the medium, which agrees with the naive expectation that medium signatures should increase for a larger medium.  In Fig.\ \ref{fig:fig3} the ratio of proton to pion production shows the expected enhancement at low energy from recombination.  This enhancement scales with the size of the medium, and is pushed out to higher energies for larger thermal parton flow velocity.

In summary, there is a strong scaling of medium signatures with the size of the medium and clear signals for thermal partons imparting flow and increasing baryon production below 10 GeV/$c$.  These trends qualitatively agree with experimental observations, though a direct comparison was not included in this study.  A tuning of MATTER and Hybrid Hadronization in small systems ($e^+ + e^-$ and $p+p$) is underway with the expectation of an experimental comparison in the near future.  This work was supported by the US National Science Foundation under award nos.\ 1550221 and 1812431.  Portions of this research were conducted with the advanced computing resources provided by Texas A\&M High Performance Research Computing.

\end{document}